\documentclass[preprint,tightenlines,aps,prd,floatfix,nofootinbib,superscriptaddress,eqsecnum]{revtex4-1}

\usepackage[mathscr]{euscript}
\usepackage{bm}
\usepackage{dcolumn}
\usepackage{graphicx}
\usepackage[pdftex]{color}
\usepackage[sort&compress]{natbib}
\usepackage[colorlinks=true,linkcolor=blue,filecolor=blue,urlcolor=blue,citecolor=blue,pdftex,plainpages=false]{hyperref}
\usepackage{amsmath,amsfonts,amssymb,mathtools}

\newcommand{\eq}[1]{Eq.~\eqref{eq:#1}}
\newcommand{\Eq}[1]{Equation \eqref{eq:#1}}
\newcommand{\eqs}[2]{Eqs.~\eqref{eq:#1} and \eqref{eq:#2}}

\def\secref#1{Sec.~\ref{sec:#1}}

\def\Secref#1{Section~\ref{sec:#1}}

\def\rcite#1{Ref.~\cite{#1}}

\newcommand{\order}{\ensuremath{\text{O}}} 

\newcommand{\ie}{\emph{i.e.},\ }
\newcommand{\eg}{\emph{e.g.},\ }

\newcommand{\chpt}{$\chi$PT}
\newcommand{\hmchpt}{HM$\chi$PT}

\newcommand{\aschpt}{HMrAS$\chi$PT}

\newcommand{\be}{\begin{equation}}
\newcommand{\ee}{\end{equation}}
\newcommand{\bea}{\begin{eqnarray}} 
\newcommand{\eea}{\end{eqnarray}}

\newcommand{\cC}{\ensuremath{\mathcal{C}}}

\newcommand{\cJ}{\ensuremath{\mathcal{J}}}

\newcommand{\cL}{\ensuremath{\mathcal{L}}}
\newcommand{\cM}{\ensuremath{\mathcal{M}}}

\definecolor{red}{rgb}{0.8,0.0,0.0}
\definecolor{green}{rgb}{0.0,0.6,0.0}
\definecolor{darkblue}{rgb}{0.0,0.1,0.7}
\definecolor{brown}{rgb}{0.6,0.1,0.0}
\definecolor{grey}{rgb}{0.6,0.6,0.6}
\definecolor{darkgreen}{rgb}{0.0, 0.545098, 0.0}
\definecolor{applegreen}{rgb}{0.55, 0.71, 0.0}
\definecolor{purple}{rgb}{0.5,0.0,0.5}
\definecolor{babypink} {rgb}{0.64, 0.44, 0.44}
\definecolor{orange}{rgb}{1.0,0.5,0.0}

\definecolor{DARKBLUE}{rgb}{0.0,0.1,0.7}

\newcommand{\strike}[1]{{}}				

\newcommand{\half}{\ensuremath{{\textstyle\frac{1}{2}}}}
\newcommand{\third}{\ensuremath{{\textstyle\frac{1}{3}}}}

\newcommand{\Sh}{\ensuremath{\widetilde{\mathcal{S}\kern-0.15em\mathit{h}}}}
\newcommand{\Ch}{\ensuremath{\widetilde{\mathcal{C}\kern-0.15em\mathit{h}}}}

\newcommand{\BE}{\begin{displaymath}}
\newcommand{\EE}{\end{displaymath}}
\newcommand{\BNE}{\begin{equation}}
\newcommand{\ENE}{\end{equation}}
\newcommand{\BEA}{\begin{eqnarray}}
\newcommand{\EEA}{\nonumber\end{eqnarray}}

\newcommand{\leftvec}{{\raise1.5ex\hbox{$\leftarrow$}\kern-1.00em}}
\newcommand{\rightvec}{{\raise1.5ex\hbox{$\rightarrow$}\kern-1.00em}}
\newcommand{\diag}{\mathop{\rm diag}}
\renewcommand{\Re}{\mathop{\rm Re}}

\newcommand{\Tr}{\mathop{\rm Tr}}

\def\negcdot{\negmedspace\cdot\negmedspace}

\newcommand{\ageom}{\ensuremath{\alpha_\text{g}}}

\newcommand{\LamQCD}{{\ensuremath{\Lambda_\text{QCD}}}}

\newcommand{\pole}{{\text{pole}}}
\newcommand{\RS}{{\text{RS}}}
\newcommand{\MRS}{{\text{MRS}}}
\newcommand{\MSbar}{{\ensuremath{\overline{\rm MS}}}}
\newcommand{\Lambdabar}{{\ensuremath{\overline{\Lambda}}}}
\newcommand{\mbar}{{\ensuremath{\overline{m}}}}

\def\rFD2014{Bazavov:2014wgs,*Bazavov:2014lja}
\def\rPDG2016{Olive:2016xmw,*Rosner:2015wva}

\begin{document}

\title{Relations between Heavy-light Meson and Quark Masses}

\author{N.~Brambilla}
\affiliation{Physik-Department, Technische Universit\"at M\"unchen, James-Franck-Stra{\ss}e 1, 85748 Garching, Germany}
\affiliation{Institute for Advanced Study, Technische Universit\"at M\"unchen, Lichtenbergstra{\ss}e 2a, 85748 Garching, Germany}

\author{J.~Komijani}
\altaffiliation[Present address:]{~School of Physics and Astronomy, University of Glasgow, Glasgow G12 8QQ, United~Kingdom.}
\affiliation{Physik-Department, Technische Universit\"at M\"unchen, James-Franck-Stra{\ss}e 1, 85748 Garching, Germany}
\affiliation{Institute for Advanced Study, Technische Universit\"at M\"unchen, Lichtenbergstra{\ss}e 2a, 85748 Garching, Germany}

\author{A.S.~Kronfeld} 
\affiliation{Institute for Advanced Study, Technische Universit\"at M\"unchen, Lichtenbergstra{\ss}e 2a, 85748 Garching, Germany}
\affiliation{Theoretical Physics Department, Fermi National Accelerator Laboratory, \\ Batavia, IL 60510, USA}

\author{A.~Vairo}
\affiliation{Physik-Department, Technische Universit\"at M\"unchen, James-Franck-Stra{\ss}e 1, 85748 Garching, Germany}

\collaboration{TUMQCD Collaboration}
\noaffiliation

\date{\today}

\begin{abstract}
The study of heavy-light meson masses should provide a way to determine renormalized quark masses and other properties of
heavy-light mesons.
In the context of lattice QCD, for example, it is possible to calculate hadronic quantities for arbitrary values of the quark masses.
In this paper, we address two aspects relating heavy-light meson masses to the quark masses.
First, we introduce a definition of the renormalized quark mass that is free of both scale dependence and renormalon ambiguities,
and discuss its relation to more familiar definitions of the quark mass.
We then show how this definition enters a merger of the descriptions of heavy-light masses in heavy-quark effective theory and in
chiral perturbation theory (\chpt).
For practical implementations of this merger, we extend the one-loop \chpt\ corrections to lattice gauge theory with heavy-light
mesons composed of staggered fermions for both quarks.
Putting everything together, we obtain a practical formula to describe all-staggered heavy-light meson masses in terms of quark
masses as well as some lattice artifacts related to staggered fermions.
In a companion paper, we use this function to analyze lattice-QCD data and extract quark masses and some matrix elements defined in
heavy-quark effective theory.
\end{abstract}


\preprint{\hspace*{-2\parskip} FERMILAB-PUB-17/490-T \hspace*{3.25in plus 1in minus 0.5in} TUM-EFT~105/17}

\maketitle


\section{Introduction}
\label{sec:Introduction}

Six of the fundamental parameters of the Standard Model are quark masses, but because of confinement they can be determined only
indirectly via their influence on hadrons.
In order to develop precise predictions of the Standard Model, it is necessary to define the quark mass in a theoretically sound
way, based on quantum field theory, and establish relations between such definitions and (nonperturbative) quantities that can be
measured in experiments or computed in lattice gauge theory.
In this and a companion paper~\cite{Bazavov:2017mbc}, we focus on the connection to lattice QCD and, in particular, on the relation
between the mass of a heavy-light hadron and the mass, $m_Q$, of the heavy quark bound inside.
Here, ``heavy'' refers to masses much larger than the scale of QCD: $m_Q\gg\Lambda_\text{QCD}$.

Properties of heavy-light hadrons can be studied with two approximate symmetries: chiral symmetry for the light quarks and
spin-flavor symmetry for the heavy quarks.
Chiral symmetry is a symmetry of the QCD Lagrangian in the limit of massless light quarks, while spin-flavor symmetry is a symmetry
of the limit of infinitely massive quarks.
These symmetries are approximate but very pertinent in suitable kinematic regions.
The small effects of symmetry breaking can be derived with effective field theories: chiral perturbation theory (\chpt) for chiral
symmetry and heavy-quark effective theory (HQET) for heavy-quark spin-flavor symmetry.

\chpt\ and HQET can be combined into heavy-meson chiral perturbation theory (\hmchpt)
\cite{Burdman:1992gh,Yan:1992gz,Wise:1992hn}, which describes the low-momentum interactions of 
pseudo-Goldstone bosons ($\pi$, $K$ and $\eta$) and hadrons containing a heavy quark.
Let us use $B_x$ to denote a generic heavy-light pseudoscalar meson composed of a light quark $x$ and a heavy antiquark $\bar Q$,
and $B^*_x$ to denote the corresponding vector meson.
The masses of the $B_x$ and $B^*_x$ mesons, to first order in $1/m_Q$ and light-quark masses, are~\cite{Wise:1992hn,Boyd:1994pa}
\begin{equation}
    M_{B^{(*)}_x} = M_0 + 2\lambda_1 B_0 m_x + 2\lambda'_1 B_0 (m_u+m_d+m_s) - d_{B^{(*)}} \frac{3\lambda_2}{2m_Q} ,
    \label{eq:Mass:Tree-level0}
\end{equation}
where $m_x$ is the light valence mass; $m_u$, $m_d$, and $m_s$ are the light (sea) quark masses; and $m_Q$ is the heavy antiquark
mass.%
\footnote{Note that all quark masses can vary in lattice-QCD simulations.} %
The physical meaning of the terms in Eq.~(\ref{eq:Mass:Tree-level0}) is as follows.
The mass $M_0$ is the spin-averaged heavy-light meson mass in the chiral limit.
The next two terms parameterize the light-quark-mass dependence; here $B_0$ enters the leading-order relation between the pion mass
and the quark masses, $m_\pi^2=B_0(m_u+m_d)$.
The last term describes the leading spin-dependent effect; here $d_B=1$ for pseudoscalar mesons and $d_{B^*}=-\third$ for vector
mesons.
In the language of \hmchpt, $\lambda_1$, $\lambda'_1$, and $\lambda_2$ are low energy constants; $\lambda_1$ and $\lambda'_1$ are
independent of the heavy-quark mass, while $\lambda_2$ has a logarithmic dependence on $m_Q$ via $\alpha_s(m_Q)$~\cite{Wise:1992hn},
as elaborated below.
These quantities can be determined by matching \hmchpt\ to the underlying theories, namely HQET and ultimately QCD.

In HQET, the mass of the $B_x$ (or $B^{*}_x$) meson is~\cite{Falk:1992wt}
\begin{equation}
    M_{B_x^{(*)}} = m_Q + \Lambdabar + \frac{\mu_\pi^2}{2m_Q} - d_{B^{(*)}}\frac{\mu_G^2(m_Q)}{2m_Q} ,
    \label{eq:mQ_2_MH}
\end{equation}
to first order in $1/m_Q$.
Each term here has a simple physical interpretation: $\bar{\Lambda}$ is the energy of the light quarks
and gluons, $\mu_\pi^2/2m_Q$ is the kinetic energy of the heavy quark in the meson's rest frame, and $d_{B^{(*)}}\mu_G^2(m_Q)/2m_Q$
corresponds to the hyperfine energy of the interaction between the heavy quark's spin with the chromomagnetic field inside the
meson.
Although each quantity corresponds to a term in the effective
Lagrangian of HQET~\cite{Kronfeld:2000ck}, for Eq.~(\ref{eq:mQ_2_MH}) to be useful in practice, $m_Q$ and
$\Lambdabar$ require a careful definition.
The principal new result of this paper is to propose a new definition with a natural interpretation in HQET, which in turn hinges on
a recent, improved understanding of the overall normalization of the renormalon~\cite{Komijani:2017vep}.
This definition is renormalon-free and scale-independent; we call it the minimal renormalon-subtracted (MRS) mass.

Note that the kinetic term is free of renormalization, as a consequence of reparametrization 
invariance~\cite{Luke:1992cs,Brambilla:2003nt,Heinonen:2012km}.
On the other hand,
\begin{equation}
    \mu_G^2(m_Q) = C_\text{cm}(m_Q) \tilde{\mu}_G^2, 
    \label{eq:mu_G2}
\end{equation}
where $C_\text{cm}(m_Q)$, which is known to three loops~\cite{Grozin:2007fh}, is the Wilson coefficient of the chromomagnetic
operator obtained by matching HQET to QCD at scale $m_Q$, and $\tilde{\mu}_G^2$ is the matrix element of the chromomagnetic
interaction operator in HQET.

The HQET quantities on the right hand side of \eq{mQ_2_MH} depend on masses of light quarks.
Moreover, the heavy-quark mass $m_Q$ depends on the sea-quark masses via two- and higher-loop diagrams.
In this work, we exploit \chpt\ to incorporate contributions from nonvanishing masses of light quarks.
If we set all light quark masses to zero in \eq{mQ_2_MH} and compare to \eq{Mass:Tree-level0}, then we find
\begin{align}
    M_0 &= \left[ m_Q + \Lambdabar + \frac{\mu_\pi^2}{2m_Q} + \order\left( m_Q^{-2} \right)
        \right]_\text{chiral limit} ,
    \label{eq:match:M0} \\
    \lambda_2 &= \left[ \frac{1}{3} \mu_G^2(m_Q) \right]_\text{chiral limit} .
    \label{eq:match:lambda2} 
\end{align}
The logarithmic dependence of $\lambda_2$ on $m_Q$ is now manifest via Eq.~(\ref{eq:mu_G2}).

For applications in lattice QCD, it is helpful to modify \chpt\ to include lattice discretization
errors, especially since lattice-QCD computations often involve an extrapolation in light-quark masses to the physical up and down
masses, and always require a continuum extrapolation in lattice spacing.
With staggered fermions, the appropriate extension of \hmchpt\ is known as heavy-meson,
rooted, all-staggered chiral perturbation theory (\aschpt)~\cite{Bernard:2013qwa}.
In this paper, we calculate one-loop corrections to heavy-light meson masses within the framework of \aschpt.
Together with the MRS mass, these corrections are needed to analyze lattice-QCD data for masses of heavy-light mesons composed of
staggered quarks, as in Ref.~\cite{Bazavov:2017mbc}.

This paper is organized as follows.
In \secref{MRS-mass}, we introduce the MRS mass.
Then, in \secref{SChPT}, we present the \aschpt\ corrections to the heavy-light meson masses.
\Secref{Synthesis} combines the MRS mass with these expressions to provide a practical way to analyze lattice-QCD data for
heavy-light meson masses, which is implemented in a companion paper~\cite{Bazavov:2017mbc} to extract the quark masses and the HQET
matrix elements $\Lambdabar$ and $\mu_\pi^2$.
Finally, in \secref{Summary}, we summarize the results and offer some outlook, both to the removal of higher renormalons and to
applications in quarkonium.

\section{Minimal-renormalon-subtracted mass}
\label{sec:MRS-mass}

Let us return to the HQET description of heavy-light meson masses:
\begin{equation}
    M_{B^{(*)}} = m_Q + \Lambdabar + \frac{\mu_\pi^2}{2m_Q} - d_{B^{(*)}}\frac{\mu_G^2(m_Q)}{2m_Q} + \cdots.
    \label{eq:mQ_2_MH-2}
\end{equation}
Clearly, the meaning of $\Lambdabar$ is connected to the choice of the renormalized mass $m_Q$,
and similar connections persist to the higher orders in the $1/m_Q$ expansion.
In the context of HQET, the most natural choice for $m_Q$ is the pole mass, because the pole mass does not depend on the ultraviolet
regulator and therefore has the same physical interpretation in QCD and HQET.
In QCD and other non-Abelian gauge theories, the pole mass is infrared finite~\cite{Kronfeld:1998di} and gauge
independent~\cite{Kronfeld:1998di,Breckenridge:1994gs} at every order in perturbation theory, but the perturbative series relating
it to a short-distance mass diverges, owing to renormalon effects~\cite{Bigi:1994em,Beneke:1994sw}.
Indeed, the leading infrared renormalon implies an intrinsic ambiguity of order~$\LamQCD$ in the definition of the pole mass.
The meson mass is unambiguous, so this ambiguity must be canceled by $\Lambdabar$, which itself is of order~\LamQCD.
Similar ambiguities from subleading infrared renormalons link the pole mass in Eq.~(\ref{eq:mQ_2_MH-2}) to the higher-order
terms~\cite{Bigi:1994em,Beneke:1994sw}, such as $\mu_\pi^2$.

Short-distance masses are free of these long-distance effects and are also candidates for~$m_Q$.
In interpreting \eq{mQ_2_MH-2}, they all leave something to be desired.
Consider choosing a mass-independent renormalization scheme, such as \MSbar, for $m_Q$.
Then, a term of order $\alpha_s m_Q$ must be absorbed into \Lambdabar, which spoils the $1/m_Q$ power counting of
HQET~\cite{Luke:1994xd}.
Threshold masses have been introduced to alleviate this problem.
Uraltsev's kinetic mass~\cite{Uraltsev:1996rd,Uraltsev:2001ih}, the renormalon-subtracted (RS)
mass~\cite{Pineda:2001zq}, and the MSR mass~\cite{Hoang:2008yj} all introduce a new factorization scale.
The 1S mass~\cite{Hoang:1999ye} is based on quarkonium and the potential-subtracted mass~\cite{Beneke:1998rk} on the heavy-quark
potential; they also introduce (at least) one more scale into the study of heavy-light hadrons.
Moreover, the heavy-quark potential is infrared divergent starting at three loops~\cite{Brambilla:1999qa}.

Here, we propose a new threshold mass that avoids introducing an additional scale.
Our proposed mass is closely related to the RS mass, and we call it the minimal renormalon-subtracted (MRS) mass.
The RS mass removes the leading-renormalon contribution from the pole mass along with a finite piece that depends on a
renormalization point, $\nu_f$, such that (ideally) $\LamQCD\ll\nu_f\ll m_Q$.
In the MRS mass, the same goal is achieved without a new~$\nu_f$.
We advocate formulating HQET using the MRS mass for the heavy quark, because it retains the advantages of the pole mass, while
circumventing its ambiguity.
(We shall also comment on variations on the MRS mass with similar advantages.) In the construction of HQET as an effective theory of
QCD, the shift from the pole to the MRS mass can be understood as an application of the notion of a residual mass
term~\cite{Falk:1992fm}.
Shifting the (pole) mass by an amount $\delta m$, $\delta m\ll m_Q$, does not disrupt the matching of HQET to
QCD~\cite{Neubert:1994wq,Luke:1994xd}.

Because the pole mass of a quark is infrared finite~\cite{Kronfeld:1998di} and gauge
independent~\cite{Kronfeld:1998di,Breckenridge:1994gs} at each order in perturbation theory, at any finite $N$ one can write
\begin{equation}
    m_\pole = \mbar\left(1+\sum_{n=0}^{N} r_n  \alpha_s^{n+1}(\mbar) + \order(\alpha_s^{N+2}) \right)  ,
    \label{eq:m-pole:expansion}
\end{equation}
where $\mbar$ denotes the $\MSbar$-renormalized mass self-consistently evaluated at the scale $\mu=\mbar$.
Here, $\alpha_s$ is the strong running coupling in a generic mass-independent scheme; the values of the coefficients $r_n$ depend on
this scheme, which is implicit but obvious in the following equations.

At large orders, the coefficients $r_n$ grow like $n!$, so the series diverges for all $\alpha_s>0$.
To cope with the divergence, one can employ the method of Borel resummation~\cite[e.g.]{{kawai2005algebraic}}
to interpret a divergent series and replace it with a better-behaved expression.
The Borel sum of a series involves an integration in the Borel plane from a base point to infinity.
The integration path is usually defined on the real axis from the origin to $+\infty$.
But, in general, it is possible to have different integration paths that cannot be deformed to each other.
Particularly, if there are singularities on the positive real axis in the Borel plane, one cannot attribute an unambiguous Borel sum
to the divergent series.

In the case of the pole mass, one finds an ambiguity of order $\LamQCD$ stemming from an infrared renormalon singularity.
Apart from the overall normalization, the full nature of this singularity, and, hence, the large-$n$ behavior of $r_n$, stems from
the running of the gauge coupling~\cite{Beneke:1994rs}.
For the normalization, expansions based on the $r_n$ in Eq.~(\ref{eq:m-pole:expansion}) have been
proposed~\cite{Pineda:2001zq,Lee:2003hh,Hoang:2008yj,Lee:2015owa,Komijani:2017vep,Hoang:2017suc,Pineda:2017uby}.
Because the leading infrared renormalon in the pole mass is closely related to an ultraviolet renormalon in the self-energy of a
static quark~\cite{Beneke:1994sw,Beneke:1994rs} and, similarly, the heavy-quark potential~\cite{Pineda:1998PhD,Hoang:1998nz},
the leading renormalon is independent of~\mbar.
Reinterpreting this observation, \rcite{Komijani:2017vep} shows that, up to an overall multiplicative factor, the large-$n$ behavior
of $r_n$ can be obtained from the recurrence relation
\begin{equation}
    a_n = 2 \left[\beta_0 n a_{n-1} + \beta_1 (n-1) a_{n-2} + \cdots + \beta_{n-1} a_0\right],\quad n\ge1,
    \label{eq:a_n:kernel}
\end{equation}
where $\beta_n$ are the coefficients of the beta function
\begin{equation}
    \beta(\alpha_s) = \frac{d\alpha_s}{d\ln\mu^2} = -\alpha_s^2\left(\beta_0 + \beta_1\alpha_s + \beta_2\alpha_s^2 + \cdots\right),
    \label{eq:beta-fcn} 
\end{equation}
capturing the running of the coupling with renormalization point~$\mu$.

The recurrence relation can be solved in closed form in the scheme with
\begin{equation}
  \beta\left(\ageom(\mu)\right) = - \frac{\beta_0\ageom^2(\mu)}{1 - (\beta_1/\beta_0) \ageom(\mu)} .
  \label{eq:beta:geometric}
\end{equation}
We shall call the scheme implied by Eq.~(\ref{eq:beta:geometric}) ``geometric'' because it corresponds to changing the scheme
of $\alpha_s$ such that $\beta_j/\beta_0=(\beta_1/\beta_0)^j$, $j\ge0$, and summing the geometric series.
We denote the coupling in this scheme $\ageom$ and comment on other schemes below.
The geometric scheme, which is very convenient in reducing the algebra, has been employed before in the literature
on renormalons~\cite[e.g.]{Brown:1992pk,Lee:1996yk}.

In this scheme, the solution of Eq.~(\ref{eq:a_n:kernel}) is~\cite[Eq.~(2.14)]{Komijani:2017vep},
\begin{equation}
    a_n = (2\beta_0)^n \frac{\Gamma(n+1+b)}{\Gamma(2+b)} a_0, \quad n\ge1,
    \label{eq:renormal:sequence:geometric}
\end{equation}
where $b=\beta_1/(2\beta_0^2)$, and $a_0$ is the initial term in the sequence.%
\footnote{Note that Eq.~(\ref{eq:renormal:sequence:geometric}) is \emph{not} valid for $n=0$.
Without this gap in the~$a_n$, the sequence would not solve Eq.~(\ref{eq:a_n:kernel}).} %
In a general scheme, the solution would include further terms that are suppressed by powers of $1/n$; see, for instance,
Ref.~\cite[Eq.~(2.5)]{Komijani:2017vep}.
In the geometric scheme, however, these corrections all vanish.

Equation~(\ref{eq:beta:geometric}) on its own does not suffice to define the geometric scheme.
In addition, one must specify the finite part dropped in the course of one-loop renormalization
from the bare to the renormalized coupling.
Equivalently, we can choose a value for $b_1$ in the relation
\begin{equation}
    \frac{1}{\ageom(\mu)} = \frac{1}{\alpha_\MSbar(\mu)} + b_1 + b_2 \alpha_\MSbar(\mu) + \cdots.
    \label{eq:alpha-b1}
\end{equation}
We choose $b_1=0$, which imposes a close relation between the geometric and \MSbar\ schemes, as we shall see in
Eq.~(\ref{eq:Lambda=Lambda}), below.
The other constants $b_n$, $n>1$, are chosen so that the scheme-dependent part of the beta function takes the geometric form,
Eq.~(\ref{eq:beta:geometric}): $b_2=\beta_2/\beta_0-(\beta_1/\beta_0)^2$, $b_3=\half[\beta_3/\beta_0-(\beta_1/\beta_0)^3]$,~etc.

The value of the overall normalization $a_0$ is needed in order to specify that $r_n\sim a_n$ as $n\to\infty$.
In the literature, it is more common to work with $R_0=a_0/(1+b)$.
(Elsewhere, $R_0$ is denoted $N_m$~\cite{Pineda:2001zq} or $N$~\cite{Beneke:1994rs,Komijani:2017vep}.)
Thus, we work with the sequence
\begin{equation}
    R_n \equiv R_0 (2\beta_0)^n \frac{\Gamma(n+1+b)}{\Gamma(1+b)}, \quad n\ge0,
    \label{eq:R_n:def}
\end{equation}
even though it is \eq{renormal:sequence:geometric}---and not Eq.~(\ref{eq:R_n:def})---that solves the recurrence relation.
Reference~\cite[Eq.~(2.22)]{Komijani:2017vep} provides a series solution for $R_0$ in the geometric scheme:
\begin{equation}
    R_0 = \sum_{k=0}^\infty r'_k \frac{\Gamma(1+b)}{\Gamma(2+k+b)} \frac{1+k}{(2\beta_0)^k},
    \label{eq:R0:series}
\end{equation}
where
\begin{equation}
    r'_k = r_k - 2\left[\beta_0 k r_{k-1} + \beta_1 (k-1) r_{k-2} + \cdots + \beta_{k-1} r_0\right]
    \label{eq:rkprime}
\end{equation}
with the $r_k$ of Eq.~(\ref{eq:m-pole:expansion}).
Note that the $r'_k$ are free of the leading pole-mass renormalon in any scheme~\cite{Komijani:2017vep}, but when evaluating $r'_k$
for Eq.~(\ref{eq:R0:series}) the geometric scheme must be used for the~$r_k$.
Because the leading renormalon cancels in the $r'_k$, they grow slowly enough such that the series in Eq.~(\ref{eq:R0:series})
converges.

To tame via Borel summation the leading-renormalon divergence on the right-hand side of \eq{m-pole:expansion},
the pole mass can be expressed as
\begin{equation}
    m_\pole = \mbar + \mbar \int_0^\infty dt\, e^{-t/\ageom(\mu)} \sum_{n=0}^\infty \left[r_n-R_n+R_n\right]\frac{t^n}{n!} ,
\end{equation}
where the $r_n$ are in the geometric scheme.
In the following, our manipulations overlook the subleading renormalons and other sources of divergence
(such as instanton contributions), taking the attitude that the method could be generalized to handle them,
and-or their practical implications are not important.
With this proviso in mind, we rewrite the pole mass as
\begin{align}
    m_\pole &= \mbar\left(1+\sum_{n=0}^{\infty} \left[r_n-R_n\right] \ageom^{n+1}(\mbar)\right) + \cJ(\mbar) ,
    \label{eq:m-pole:expansion-2} \\
    \cJ(\mu) &\equiv \mu \int_0^\infty dt \,e^{-t/\ageom(\mu)} \sum_{n=0}^\infty \frac{R_n}{n!}t^n.
    \label{eq:J:def}
\end{align}
In \eq{m-pole:expansion-2}, the power series in $\ageom$ with $[r_n-R_n]$ is better-behaved than
that in \eq{m-pole:expansion}, because it is free of the leading infrared renormalon~\cite{Pineda:2001zq}.

Let us now examine the second term in \eq{m-pole:expansion-2},
$\cJ(\mu)$, which contains the leading infrared renormalon.
Plugging \eq{R_n:def} into the series inside the integral, one finds
\begin{equation}
    \sum_{n=0}^{\infty} \frac{\Gamma(n+1+b)}{\Gamma(1+b) n!}z^n = \frac{1}{(1-z)^{1+b}} ,
\end{equation}
where $z=2\beta_0 t$.
Here, following the convention in performing a Borel sum, a power series is summed
in its region of validity, $|z|<1$, and then extended by analytic continuation.
Thus,
\begin{equation}
    \cJ(\mu) = \frac{R_0}{2\beta_0} \mu \int_0^\infty dz\, \frac{e^{-z/(2\beta_0\ageom(\mu))}}{(1-z)^{1+b}}.
    \label{eq:J:0}
\end{equation}
A full definition must specify the path of integration in the complex $z$ plane
and the branch cut for the multivalued function $(1-z)^{-(1+b)}$.

Our strategy is to divide the integral into two pieces:
\begin{equation}
    \cJ(\mu) = \cJ_\MRS(\mu) + \delta m,
    \label{eq:J:split}
\end{equation}
where
\begin{align}
    \cJ_\MRS(\mu) &= \frac{R_0}{2\beta_0} \mu \int_0^1      dz\,\frac{e^{-z/[2\beta_0 \ageom(\mu)]}}{(1-z)^{1+b}} ,
    \label{eq:JMRS:def} \\
    \delta m      &= \frac{R_0}{2\beta_0} \mu \int_1^\infty dz\,\frac{e^{-z/[2\beta_0 \ageom(\mu)]}}{(1-z)^{1+b}} .
    \label{eq:deltam:def}
\end{align}
Because the integration in $\cJ_\MRS(\mu)$ lies within the unit disk, it does not depend on the contour.
In both pieces, the integral is defined first for $\Re b<0$ and by analytic continuation elsewhere.
As discussed below, the first term, $\cJ_\MRS$, is unambiguous; the full ambiguity of the leading renormalon is contained
in~$\delta m$.

In $\delta m$, the explicit $\mu$ dependence and implicit dependence via $\ageom(\mu)$ cancel.
This feature can be shown, and the branch cut moved into a prefactor, by setting $z=1+2\beta_0\ageom(\mu)x$, yielding
\begin{align}
    \delta m &= (-1)^{1+b} \frac{R_0}{2\beta_0} \mu \frac{e^{-1/[2\beta_0\ageom(\mu)]}}{[2\beta_0\ageom(\mu)]^b}
        \int_0^\infty dx\, x^{-b-1}e^{-x} \nonumber \\
        &= -(-1)^{b} \frac{R_0}{2^{1+b}\beta_0} \Gamma(-b) \Lambda_\text{g},
    \label{eq:deltam:Lambdag}
\end{align}
where $(-1)^b=e^{\pm i\pi b}$ if the contour is set slightly above ($+$) or below ($-$) the positive real axis.
Remarkably, the conventional~\cite{Bardeen:1978yd,Celmaster:1979km,Chetyrkin:2000yt} constant of integration introduced when solving
Eq.~(\ref{eq:beta:geometric}) appears:
\begin{equation}
    \Lambda_\text{g} = \mu \frac{e^{-1/[2\beta_0\ageom(\mu)]}}{[\beta_0\ageom(\mu)]^b}.
    \label{eq:Lambda:geometric}
\end{equation}
This constant is independent of $\mu$ to all orders, so, as announced, $\delta m$ does not depend on~$\mu$.
Moreover, owing to the choice $b_1=0$ in Eq.~(\ref{eq:alpha-b1}), it also follows, to all
orders~\cite{Bardeen:1978yd,Celmaster:1979km}, that
\begin{equation}
    \Lambda_\text{g} = \Lambda_\MSbar.
    \label{eq:Lambda=Lambda}
\end{equation}
Finally, the overall normalization of the renormalon, $R_0$, is the same in both schemes~\cite{Beneke:1998ui}, and indeed in any
scheme such that $b_1=0$.
%
\nocite{Pineda:2001zq,Brambilla:2009bi,Brambilla:2010pp}
\phantom{\cite{Bazavov:2012ka,*Bazavov:2014soa}}
\nocite{Pineda:2013lta,Ayala:2014yxa,Berwein:2015vca}

At this point, we can present the main result of this section, namely a new proposal for the mass to be used in setting up HQET 
and, thus, Eq.~(\ref{eq:mQ_2_MH-2}).
Reference~\cite{Pineda:2001zq} introduces the \emph{renormalon-subtracted} (RS) mass, which in our notation can be written%
\begin{align}
    m_\RS(\nu_f) \equiv m_\pole - \cJ(\nu_f),
    \label{eq:mpole2mRS}
\end{align}
clearly removing the renormalon ambiguity.%
\footnote{For some practical applications of the RS mass, see Ref.~\cite{Pineda:2001zq,Bali:2003jq,Brambilla:2009bi,%
Brambilla:2010pp,Bazavov:2012ka,Pineda:2013lta,Ayala:2014yxa,Berwein:2015vca}.}
Our proposal is to subtract the minimal piece of $\cJ(\mu)$ that removes the renormalon, namely $\delta m$:
\begin{align}
    m_\MRS &\equiv m_\pole - \delta m ,
    \label{eq:mMRS=mpole-dm} \\
        &= \mbar\left(1+\sum_{n=0}^{\infty} \left[r_n-R_n\right] \ageom^{n+1}(\mbar)\right) + \cJ_\MRS(\mbar).
    \label{eq:mSI2mMRS}
\end{align}
We call $m_\MRS$ the \emph{minimal renormalon-subtracted} (MRS) mass, because it removes only the part of $\cJ(\mu)$ that is
ambiguous.
Equation~(\ref{eq:mMRS=mpole-dm}) should be considered a formal, intuitive definition (both terms are ambiguous), while
Eq.~(\ref{eq:mSI2mMRS}) gives a rigorous (apart from subleading renormalons and similar obstacles, overlooked in this analysis) and,
thus, operational definition.
As the notation suggests, and as Eq.~(\ref{eq:deltam:Lambdag}) shows, the difference in using $m_\MRS$ instead of $m_\pole$ amounts
to a shift by an amount $\delta m\sim\Lambda_\MSbar\ll m_Q$, which does not disrupt the $1/m_Q$
expansion~\cite{Neubert:1994wq,Luke:1994xd}.

To evaluate $m_\MRS$, one can proceed as follows.
First, one can carry out the integration in Eq.~(\ref{eq:JMRS:def}), obtaining
\begin{equation}
    \cJ_\MRS(\mbar) = \frac{R_0}{2\beta_0} \mbar e^{-1/[2\beta_0\ageom(\mbar)]}
        \Gamma(-b) \gamma^\star\left(-b,-[2\beta_0\ageom(\mbar)]^{-1}\right) ,
  \label{eq:J-MRS:def:Incomplete-gamma}
\end{equation}
where $\gamma^\star(a,w)\equiv[w^{-a}/\Gamma(a)]\int_0^wdt\,t^{a-1}e^{-t}$ is the limiting function of the incomplete gamma
function~\cite{Davis:GammaFunction}.
In practice, the power series
\begin{equation}
    \Gamma(-b) \gamma^\star(-b,-y) = \sum_{n=0}^\infty \frac{y^n}{n!(n-b)} 
    \label{eq:JMRS:series}
\end{equation}
converges quickly.
This series can be derived from \eq{JMRS:def} by changing variables to $x=1-z$, expanding $e^{x/y}$ in powers of $x$, and
integrating term-by-term.
Next, the sum in Eq.~(\ref{eq:mSI2mMRS}) is carried out to the order available, taking care to be consistent with the scheme.
The $r_n$ are used directly in Eq.~(\ref{eq:mSI2mMRS}) and also in the $R_n$ via $R_0$ [Eq.~(\ref{eq:R0:series})].
Because the sequence corresponding to the leading infrared renormalon has been removed, the power series for the MRS mass is
better-behaved than the uninterpreted series, \eq{m-pole:expansion}.
As shown in Sec.~\ref{sec:Synthesis}, Eqs.~(\ref{eq:rn3})--(\ref{mSI2MRS:sequence}), the better behavior is realized in
practice.

This concludes the derivation and implementation of the MRS scheme for the quark mass.
We end this section with a few interesting and useful remarks.

It is interesting to examine the asymptotic expansions for $\cJ_\MRS$ and $m_\MRS$,
because they underscore a key feature of asymptotic expansions: the same expansion can hold for more than one quantity.
As a function of~$\ageom$, $\cJ_\MRS(\mbar)$ has a convergent expansion in powers of~$1/\ageom(\mbar)$,
but it is not an analytic function about $\ageom(\mbar)=0$.
Its power expansion leads to the asymptotic expansion
\begin{equation}
    \cJ_\MRS(\mbar) \sim \mbar \sum_{n=0}^{\infty} R_n  \ageom^{n+1}(\mbar), \quad \text{as } \ageom(\mbar)\to 0.
    \label{eq:J-MRS-asymptotic}
\end{equation}
The ambiguous function $\cJ(\mbar)$, given in \eq{J:def}, has the same asymptotic expansion as $\cJ_\MRS(\mbar)$, because its
ambiguous component $\delta m$ is exponentially small in $1/\ageom(\mbar)$ and, thus, not visible to conventional asymptotics.
(See \rcite{Boyd1999} for a discussion on exponentially small terms in asymptotic calculations.) %
Similarly, putting \eqs{mSI2mMRS}{J-MRS-asymptotic} together for the MRS~mass, one finds
\begin{equation}
    m_\MRS \sim \mbar + \mbar \sum_{n=0}^{\infty} r_n \ageom^{n+1}(\mbar),
        \quad \text{as } \ageom(\mbar)\to 0,
    \label{eq:m-MRS-asymptotic}
\end{equation}
which is the same series as the asymptotic expansion of the pole mass.
Note that with suitable changes to the $R_n$ and $r_n$,
Eqs.~(\ref{eq:J-MRS-asymptotic}) and~(\ref{eq:m-MRS-asymptotic}) may be extended to any renormalization scheme.

Another method to fix the ambiguity of the pole mass is the so-called principal value (PV)
prescription~\cite[e.g.]{Beneke:1994rs,Lee:2003hh,Bali:2003jq}, which removes the ambiguous imaginary part of $m_\pole$:
\begin{align}
   m_\text{PV} &\equiv \Re m_\pole
   \label{eq:mMRS2mPV:def} \\ 
        &= m_\MRS - \cos(\pi b) \frac{R_0}{2^{1+b}\beta_0} \Gamma(-b) \Lambda_\text{g} .
  \label{eq:mMRS2mPV}
\end{align}
Like $m_\MRS$, $m_\text{PV}$ also contains no subtraction scale.
The PV mass has the advantage that it remains finite when $b$ tends to a positive integer or zero.
We prefer the MRS scheme, because the PV scheme entails an arbitrary decision to replace the multivalued factor $(-1)^b$ with its
real part.
In the MRS mass, on the other hand, all ambiguity is swept into $\delta m$ and absorbed into the residual mass of HQET, as discussed
below.

As just implied, the MRS mass is not ideal when $b$ tends to a non-negative integer, in which case $\delta m$ is no longer of order
$\Lambda_\text{g}$.
Suppose, for example, $b\to0$.
Then, we can use a variant MRS$'$ such that:
\begin{align}
    m_{\MRS'}          &= m_\pole - \delta m', \\
    \delta m'          &= \delta m        - \frac{R_0}{2\beta_0 b} \Lambda_\text{g}, \\
    \cJ_{\MRS'}(\mbar) &= \cJ_\MRS(\mbar) + \frac{R_0}{2\beta_0 b} \Lambda_\text{g},
    \label{eq:NoNo}
\end{align}
which introduces a term chosen to cancel the $n=0$ term in Eq.~(\ref{eq:JMRS:series}) as $b\to 0$.
Other variants could remove the singular term in Eq.~(\ref{eq:JMRS:series}) as $b$ tends to any positive integer.

With the MRS mass for heavy quarks, we can now revisit the discussion in \secref{Introduction} on resolving the problem of
ambiguities in HQET.
We reconstruct the HQET Lagrangian using the MRS mass for the heavy quark and setting the corresponding residual mass%
\footnote{Our residual mass $\delta m$ is minus the residual mass introduced in \rcite{Falk:1992fm}.}
to the ambiguous term $\delta m$. Then, $M_0$ in \eq{match:M0} reads
\begin{align}
    M_0 &= \left[ m_\MRS + \delta m + \Lambdabar + \frac{\mu_\pi^2}{2m_\MRS} +
        \order\left( m_\MRS^{-2} \right)\right]_\text{chiral-limit},
        \nonumber \\
        &= \left[ m_\MRS + \Lambdabar_\MRS +       \frac{\mu_\pi^2}{2m_\MRS} +
        \order\left( m_\MRS^{-2} \right)\right]_\text{chiral-limit},
    \label{eq:match:M0:MRS}
\end{align}
where $\Lambdabar_\MRS = \Lambdabar + \delta m$ is a well-defined quantity, because it is associated with two well-defined
quantities, $M_0$ and~$m_\MRS$.
We do not attribute the MRS scheme to $\mu_\pi^2$ in \eq{match:M0:MRS}, because, as defined here, the MRS scheme does not
subtract subleading renormalons in the pole mass, such as the one canceled by~$\mu_\pi^2$.
Renormalon-free definitions of \Lambdabar\ in other schemes, such as PV and MRS$'$, follow by analogy with
Eq.~(\ref{eq:match:M0:MRS}).

Finally, it is worth recording the relations between the RS and MRS schemes:%
\footnote{\Eq{mMRS2mRS} provides an alternative way to calculate the RS mass, similar to that used in \rcite{Bali:2003jq}.}
\begin{align}
    m_\RS(\nu_f) &= m_\MRS - \cJ_\MRS(\nu_f),
    \label{eq:mMRS2mRS} \\
    \Lambdabar_\RS(\nu_f) &= \Lambdabar_\MRS + \cJ_\MRS(\nu_f),
    \label{eq:Lambbdabar:MRS2RS}
\end{align}
which follow immediately from Eqs.~(\ref{eq:mpole2mRS}) and~(\ref{eq:mMRS=mpole-dm}), with $\cJ_\MRS(\nu_f)$ evaluated as described 
in and below Eq.~(\ref{eq:J-MRS:def:Incomplete-gamma}).

\section{Heavy-light meson masses in staggered \boldmath\texorpdfstring{\chpt}{ChPT}}
\label{sec:SChPT}

In this section, we discuss the corrections to heavy-light meson masses from the ``pion cloud,'' using heavy-meson rooted
all-staggered chiral perturbation theory (\aschpt), which was developed to compute these corrections for heavy-light decay
constants~\cite{Bernard:2013qwa}.
\aschpt\ incorporates aspects of staggered fermions into standard \hmchpt\
\cite{Burdman:1992gh,Yan:1992gz,Wise:1992hn,Boyd:1994pa}, in particular the extra pseudo-Goldstone bosons arising from the four
fermion species, or tastes, that emerge from each staggered fermion field in the continuum limit.
For the sea quarks, \aschpt\ also accounts for taking the fourth root of the staggered determinant,
which is done to simulate one flavor per staggered fermion field.

The light pseudoscalar sector of the leading order \aschpt\ Lagrangian includes terms to describe the breaking of SU(4) taste
symmetry at order~$a^2$~\cite{Lee:1999zxa}.
For example, a light pseudoscalar meson with flavors $x$ and $y$ and mesonic taste $\Xi$ has, at the tree 
level~\cite[Eq.~(18)]{Aubin:2003mg},
\begin{equation}
    m^2_{{xy}_\Xi} = B_0 (m_x + m_y) + a^2 \Delta_\Xi
    \label{eq:pion-mass-sq:Xi}
\end{equation}
with $\Delta_P=0$, where $P$ is the so-called pseudoscalar taste.
(The full set of 16 mesonic tastes is given the labels $\{I,V,T,A,P\}$, corresponding to multiplets of 
$(1,4,6,4,1)$, respectively~\cite{Kronfeld:2007ek}.)
In contrast, the heavy-light sector of the leading-order \aschpt\ Lagrangian---and also the one-loop heavy-light meson
propagator---are taste invariant~\cite[Eq.~(27) and~(151)]{Bernard:2013qwa}.
Therefore, all taste violations in the heavy-light meson masses at next-to-leading order (NLO) come from the NLO terms in the
HMrAS\chpt\ Lagrangian, treated at the tree level; see, for instance, Eqs.~(125) and~(127) and Tables~I and II in
Ref.~\cite{Bernard:2013qwa}.
Because these terms can be added straightforwardly, we drop the taste index of heavy-light mesons, for convenience, and we neglect
the tree-level lattice corrections to their masses, which can be restored at the end, as required.

The leading corrections to heavy-light meson masses come from two sources: a one-loop contribution from the self energy of the
heavy-light meson, and the tree-level contributions from the parts of the \aschpt\ Lagrangian that explicitly break the chiral
symmetry of the light quarks and the spin-flavor symmetry of the heavy quarks.
The corresponding terms in the effective Lagrangian are
\begin{align}
    \cL_0 &= M_0\Tr\left(\overline{H} H\right),
    \label{eq:LM0} \\
    \cL_1 &=  -i\Tr\left(\overline{H} H v\negcdot \leftvec{D}\right) +
        g_\pi   \Tr\left(\overline{H}H\gamma^{\mu}\gamma_5 \mathbb{A}_{\mu}\right),
    \label{eq:L1} \\
    \cL_m &= 2\lambda_1 B_0 \Tr\left(\overline{H} H\cM^+\right) +
        2\lambda'_1 B_0 \Tr\left(\overline{H} H\right)\Tr\left(\cM^+\right) -
        \frac{\lambda_2}{4m_Q} \Tr\left(\overline{H}\sigma^{\mu\nu} H \sigma_{\mu\nu}\right),
    \label{eq:L_mass}
\end{align}
where $M_0$ from Eq.~(\ref{eq:match:M0}) is included for convenience.%
\footnote{One can choose the generators of heavy-quark flavor symmetry to be compatible with a nontrivial heavy-quark mass matrix 
$m_Q$~\cite[Sec.~III]{Kronfeld:2000ck}.
Also, the $\mu_\pi^2$ term could, if desired, be moved from $\cL_0$ to $\cL_m$.}
Here $H$ is a heavy-meson field for pseudoscalar and vector mesons,
$\mathcal{M}^+=\half[\sigma\mathcal{M}\sigma+\sigma^\dagger\mathcal{M}\sigma^\dagger]$ with $\mathcal{M}=\diag(m_u,m_d,m_s)$,
$\sigma=\exp(i\Phi/2f)$, $\sigma_{\mu\nu}=(i/2)[\gamma_\mu,\gamma_\nu]$,
$v^\mu$ is the four-velocity of HQET,
$D_\mu=\partial_\mu-i\mathbb{V}_\mu$,
$\mathbb{V}_\mu=\frac{i}{2}(\sigma^\dagger\partial_\mu\sigma+\sigma\partial_\mu\sigma^\dagger)$,
$\mathbb{A}_\mu=\frac{i}{2}(\sigma^\dagger\partial_\mu\sigma-\sigma\partial_\mu\sigma^\dagger)$,
and $g_\pi$ is the $H$-$H^*$-$\pi$ coupling.
The matrix field~$\Phi$ contains the $16n^2-1$ pseudo-Goldstone bosons for $n$ staggered light flavors, and $f$ is the decay
constant of these bosons (in the chiral limit).
As discussed elsewhere~\cite{Aubin:2003mg}, $\Phi$ also contains the flavor-taste singlet $\eta'_I$,
which is removed by taking $m_{\eta'_I}$ to infinity, relative to the pseudo-Goldstone boson masses,
once these contributions have been isolated.
For full details of the notation, see Refs.~\cite{Aubin:2003mg,Aubin:2005aq,Bernard:2013qwa}.
In Eq.~(\ref{eq:L_mass}), we follow the normalization of \rcite{Arndt:2004bg} for $\lambda_1$ and $\lambda'_1$
and of \rcite{Stewart:1998ke} for~$\lambda_2$, but with opposite sign so that our $\lambda_2>0$.
Further, in Eq.~(\ref{eq:L_mass}), we use the power counting adopted in \rcite{Stewart:1998ke}; $m_q\sim1/m_Q$ with $m_q$
being a generic light quark mass.

The heavy-light mesons carry both light flavor and taste indices.
As mentioned above, the taste index is unnecessary at this order, so we write $B_{x}$ and $B^*_{x}$ for pseudoscalar and vector
mesons, respectively, with light valence flavor $x$.
Their masses are then
\begin{subequations}
    \label{eq:Mass-B-all}
    \begin{align}
        M_{B_x}   &= M_0 + 2\lambda_1 B_0 m_x + 2\lambda'_1 B_0\Tr \cM - \frac{3\lambda_2}{2m_Q} + \delta M_{B_x},
        \label{eq:Mass-B} \\
        M_{B^*_x} &= M_0 + 2\lambda_1 B_0 m_x + 2\lambda'_1 B_0\Tr \cM +  \frac{\lambda_2}{2m_Q} + \delta M_{B^*_x},
        \label{eq:Mass-B*}
    \end{align}
\end{subequations}
where the three terms after $M_0$ are the tree-level corrections, which are immediate from Eq.~(\ref{eq:L_mass}), and
$\delta{}M_{B^{(*)}_x}$ is the one-loop correction from the pion cloud.

To obtain $\delta M_{B^{(*)}_x}$, one must evaluate the one-loop self energy on shell.
Although the flavor and hyperfine splittings in the heavy-meson propagator are formally subleading in~$m_q$ and~$1/m_Q$,
respectively, it is quantitatively sensible to include them.
In practice, both the flavor splittings (\eg $M_{D_s}-M_D\approx99$~MeV) and hyperfine splittings (\eg $M_{D^*}-M_D\approx144$~MeV)
are not much different than the pion mass, so they influence the behavior of the nonanalytic terms in the one-loop contribution.

The self energy in \aschpt\ has already been computed in Ref.~\cite{Bernard:2013qwa},
which in turn relies on Ref.~\cite{Aubin:2005aq}.
There the self energy is used to obtain the wave-function renormalization needed to compute the one-loop
corrections to the heavy-light decay constants.
(Some steps are not spelled out in Refs.~\cite{Bernard:2013qwa,Aubin:2005aq}, so we rederived the self energy from scratch.) %
Let us first consider the case in which the three light sea quarks have distinct masses, \pagebreak \ie $1+1+1$.
We find
\begin{align}
    \delta M_{B_x}\Bigr|_{1+1+1}   &= - \frac{3g_\pi^2}{16\pi^2f^2} \Biggl\{
        \frac{1}{16}\sum_{\mathscr{S},\Xi} K_1(m_{\mathscr{S}x_{\Xi}},\Delta^* + \delta_{\mathscr{S}x})
    \label{eq:dMB} \\
        & \qquad + \frac{1}{3} \sum_{j\in \cM_I^{(3,x)}} \frac{\partial}{\partial m^2_{X_I}}
            \left[R^{[3,3]}_{j}(\cM_I^{(3,x)}; \mu^{(3)}_I) K_1(m_{j},    \Delta^*) \right] \nonumber \\
        & \qquad + \biggl( a^2\delta'_V \sum_{j\in \cM_V^{(4,x)}} \frac{\partial}{\partial m^2_{X_V}}
            \left[R^{[4,3]}_{j}(\cM_V^{(4,x)}; \mu^{(3)}_V) K_1(m_{j},    \Delta^*) \right]
		  + [V\to A] \biggr)  
	    \Biggr\},
    \nonumber \\[1em]
    \delta M_{B^*_x}\Bigr|_{1+1+1} &= -  \frac{g_\pi^2}{16\pi^2f^2} \Biggl\{
        \frac{1}{16}\sum_{\mathscr{S},\Xi} \Bigl(K_1(m_{\mathscr{S}x_{\Xi}},-\Delta^* + \delta_{\mathscr{S}x})
            + 2 K_1(m_{\mathscr{S}x_{\Xi}},\delta_{\mathscr{S}x})\Bigr)
    \label{eq:dMB*} \\
        & \qquad + \frac{1}{3} \sum_{j\in \cM_I^{(3,x)}} \frac{\partial}{\partial m^2_{X_I}}
            \left[R^{[3,3]}_{j}(\cM_I^{(3,x)}; \mu^{(3)}_I) K_1^*(m_{j}, -\Delta^*) \right]
    \nonumber \\
        & \qquad + \biggl( a^2\delta'_V \sum_{j\in \cM_V^{(4,x)}}\frac{\partial}{\partial m^2_{X_V}}
            \left[R^{[4,3]}_{j}(\cM_V^{(4,x)}; \mu^{(3)}_V) K_1^*(m_{j}, -\Delta^*) \right] + [V\to A]\biggr)  
	    \Biggr\} , \nonumber
\end{align}
where $\mathscr{S}$ runs over the three light sea flavors ($u$, $d$, $s$); $\Xi$ runs over 16 meson tastes;
$m_{\mathscr{S}x_\Xi}$ and $m_{X_\Xi}$ are masses of mesons with taste $\Xi$ and, respectively, flavor $\bar{\mathscr{S}}x$
and~$\bar{x}x$; and $\delta'_V$ ($\delta'_A$) is the (axial) vector hairpin low-energy constant.

The notation for the loop functions is
\begin{align}
    K_1(m,\Delta)   &= \Delta J_1(m,\Delta), \label{eq:K1:def} \\
    K_1^*(m,\Delta) &= K_1(m,\Delta) + 2K_1(m,0) = K_1(m,\Delta) + \frac{4\pi}{3}m^3, \\
    J_1(m,\Delta)   &= \left(\frac{2}{3}\Delta^2-m^2\right) \ln\left(\frac{m^2}{\Lambda_\chi^2}\right) 
        + \frac{4}{3}  \left(\Delta^2-m^2\right) F\left(\frac{m}{\Delta}\right)
        - \frac{10}{9} \Delta^2
        + \frac{4}{3}  m^2 ,
    \label{eq:J1def} \\
    F(1/x)          &=
        \begin{cases} \displaystyle
             -\frac{\sqrt{1-x^2}}{x_{\phantom{g}}}\left(\frac{\pi}{2} - \tan^{-1}\frac{x}
            {\sqrt{1-x^2}}\right), & \text{if $ |x|\le 1$}, \\ \displaystyle
            \frac{\sqrt{x^2-1}}{x}\ln\left(x + \sqrt{x^2-1}\right), & \text{if $|x|\ge 1$}.
        \end{cases}
    \label{eq:Fdef}
\end{align}
In Eq.~(\ref{eq:J1def}), $\Lambda_\chi$ is the renormalization scale of~\chpt.
The residue symbols originating from the hairpin diagrams take two sets of masses as arguments:
\begin{equation}
    R_j^{[n,k]}(\{m\}; \{\mu\}) = \frac{\prod_{i=1}^k(\mu^2_i- m^2_j)}{\prod_{r=1|r\neq j}^n(m^2_r - m^2_j)}.
    \label{eq:Rjnk}
\end{equation}
The sets needed in Eqs.~(\ref{eq:dMB}) and~(\ref{eq:dMB*}) are
\begin{align}
    \mu^{(3)}_\Xi   &= \{m_{U_\Xi}, m_{D_\Xi},     m_{S_\Xi}\}, \text{ for } \Xi \in \{I, V, A\},
    \label{eq:mu3Xi} \\
    \cM^{(3,x)}_I   &= \{m_{X_I},   m_{\pi^0_I},   m_{\eta_I}  \}, \\
    \cM^{(4,x)}_\Xi &= \{m_{X_\Xi}, m_{\pi^0_\Xi}, m_{\eta_\Xi}, m_{\eta'_\Xi}\}, \text{ for } \Xi \in \{V, A\}.
\end{align}
Here, $U_\Xi$, $D_\Xi$, and $S_\Xi$ denote $\bar{u}u_\Xi$,
$\bar{d}d_\Xi$, and $\bar{s}s_\Xi$ mesons, respectively, while $\pi^0_\Xi$, $\eta_\Xi$, and
$\eta'_\Xi$ denote the eigenstates after mixing among neutral mesons is taken into account.
Only the flavor-taste singlet $\eta'_I$ receives a large mass related to the axial anomaly and,
as mentioned above, has been removed.

Finally, we have the hyperfine and flavor splittings.
$\Delta^*=2\lambda_2/m_Q$ is the lowest-order hyperfine splitting, and $\delta_{\mathscr{S}x}$ is
the lowest-order flavor splitting between heavy-light mesons with light quarks of flavor $\mathscr{S}$ and $x$.
$\delta_{\mathscr{S}x}$ can be written in terms of pion masses as
\begin{equation}
    \delta_{\mathscr{S}x} = 2 \lambda_1 B_0 (m_\mathscr{S}-m_x) \approx \lambda_1 (m^2_{\mathscr{SS}_P} - m^2_{X_P}),
    \label{eq:delta_sx-ask}
\end{equation}
where $P$ again denotes pseudoscalar taste.

In lattice-QCD simulations, the up and down sea-quark masses are often taken equal, which is then denoted as a ($2+1$)-flavor sea.
Throughout this paper, we write $m_l \equiv \frac{1}{2}(m_u+m_d)$.
For the $2+1$ case, we find
\begin{align}
    \delta M_{B_x}\Bigr|_{2+1}   &= - \frac{3g_\pi^2}{16\pi^2f^2} \Biggl\{
        \frac{1}{16}\sum_{\mathscr{S},\Xi} K_1(m_{\mathscr{S}x_{\Xi}},\Delta^* + \delta_{\mathscr{S}x})
    \label{eq:dMB-2+1} \\
        & \qquad + \frac{1}{3}\sum_{j\in \cM_I^{(2,x)}} \frac{\partial}{\partial m^2_{X_I}}
            \left[R^{[2,2]}_{j}(\cM_I^{(2,x)};     \mu^{(2)}_I) K_1(m_{j},   \Delta^*) \right]
    \nonumber \\
        & \qquad + \biggl( a^2\delta'_V \sum_{j\in \hat{\cM}_V^{(3,x)}} \frac{\partial}{\partial m^2_{X_V}}
            \left[R^{[3,2]}_{j}(\hat{\cM}_V^{(3,x)}; \mu^{(2)}_V) K_1(m_{j},   \Delta^*) \right] + [V\to A]\biggr)  
	    \Biggr\},
    \nonumber \\[1em]
    \delta M_{B^*_x}\Bigr|_{2+1} &= -  \frac{g_\pi^2}{16\pi^2f^2} \Biggl\{
	    \frac{1}{16}\sum_{\mathscr{S},\Xi} \Bigl(K_1(m_{\mathscr{S}x_{\Xi}},-\Delta^* + \delta_{\mathscr{S}x})
            + 2 K_1(m_{\mathscr{S}x_{\Xi}},\delta_{\mathscr{S}x})\Bigr)
    \label{eq:dMB*-2+1} \\
        & \qquad + \frac{1}{3} \sum_{j\in \cM_I^{(2,x)}} \frac{\partial}{\partial m^2_{X_I}}
            \left[R^{[2,2]}_{j}(\cM_I^{(2,x)};    \mu^{(2)}_I) K_1^*(m_{j}, -\Delta^*) \right]
    \nonumber \\
        & \qquad + \biggl( a^2\delta'_V \sum_{j\in \hat{\cM}_V^{(3,x)}} \frac{\partial}{\partial m^2_{X_V}}
            \left[R^{[3,2]}_{j}(\hat{\cM}_V^{(3,x)}; \mu^{(2)}_V) K_1^*(m_{j}, -\Delta^*) \right] + [V\to A]\biggr)  
	    \Biggr\},
    \nonumber
\end{align}
with sets
\begin{align}
    \mu^{(2)}_\Xi   &= \{m_{L_\Xi}, m_{S_\Xi}\}, \text{ for } \Xi \in \{I, V, A\}, \\
    \cM^{(2,x)}_I   &= \{m_{X_I},   m_{\eta_I}  \}, \\
    \hat{\cM}^{(3,x)}_\Xi &= \{m_{X_\Xi}, m_{\eta_\Xi}, m_{\eta'_\Xi}\}, \text{ for } \Xi \in \{V, A\},
\end{align}
where $L_\Xi$, analogously to Eq.~(\ref{eq:mu3Xi}), denotes an $\bar{l}l_\Xi$ meson.
In the continuum limit of full QCD (where the light valence quark is one of the sea quarks), the nonanalytic terms in these
expressions agree with the corresponding results, Eqs.~(A6), (A7), (A8) and~(A9), in HM$\chi$PT~\cite{Mehen:2005hc}; there are
unimportant differences in the analytic terms, stemming from a different choice of counterterms.

As they stand, the expressions for $\delta M_{B^{(*)}_x}$ are not completely independent of the chiral scale 
$\Lambda_\chi$, but this dependence can be absorbed into counterterms.
Most of these are the same as in continuum \hmchpt, and they are formally of higher order in the \chpt\ expansion.
An interesting example is suppressed by $a^2$ and can be handled via
\begin{equation}
    \delta M_{B^{(*)}_x} \to \delta M_{B^{(*)}_x} + \lambda_{a^2} a^2
        \frac{3g_\pi^2}{16\pi^2f^2} \left[ \bar{\Delta} \sum_\mathscr{S}\delta_{\mathscr{S}x} 
        + d_{B^{(*)}}\Delta^* \left( 3\bar{\Delta} -\third\Delta_I + \delta'_V + \delta'_A\right)\right],
    \label{eq:barDelta-counterterm}
\end{equation}
where $d_{B^{(*)}}=1(-\third)$, and the nearly%
\footnote{These splittings are empirically nearly flavor independent (see, \eg Ref.~\cite[Fig.~6]{Bazavov:2012xda}),
which can be understood because the flavor dependence vanishes at leading order in staggered \chpt~\cite{Aubin:2003mg}.}
$\mathscr{S}x$-independent quantities
\begin{align}
    a^2\Delta_I &= m_{\mathscr{S}x_I}^2 - m_{\mathscr{S}x_P}^2, \\
    a^2\bar{\Delta} &= \frac{1}{16}\sum_{\Xi}\left( m_{\mathscr{S}x_\Xi}^2 - m_{\mathscr{S}x_P}^2\right)
    \label{eq:barDelta}
\end{align}
stem from the taste splittings in the logarithmic terms in $K_1$; cf.\ Eqs.~(\ref{eq:K1:def}) and~(\ref{eq:J1def}).
The remaining $\Lambda_\chi$ dependence can be absorbed into $\lambda_1$, $\lambda'_1$ and higher order counterterms in continuum
\hmchpt.

%
%
%

Finite-volume effects can be incorporated into $\delta M_{\raisebox{5pt}{$\scriptstyle B^{(*)}_x$}}$ by substituting
$K_1(m,\Delta)$ with its finite-volume
counterpart.
Equation~(92) of Ref.~\cite{Aubin:2007mc} gives the finite-volume version of $J_1(m,\Delta)$.
With \eq{K1:def}, we then find, for a spatial volume $L^3$,
\begin{equation}
    K_1(m,\Delta) \to K_{1,\text{FV}}(m,\Delta) = K_1(m,\Delta) + \delta K_1(m,\Delta,L),
    \label{eq:K1fv-def}
\end{equation}
where 
\begin{equation}
    \delta K_1(m,\Delta,L) = -\frac{m^2}{3}\Delta\delta_1(mL) - 16\pi^2 \frac{m^2-\Delta^2}{3} J_\text{FV}(m,\Delta,L),
  \label{eq:K1fv-only-def}
\end{equation}
with $\delta_1(mL)$ and $J_\text{FV}(m,\Delta,L)$ given, respectively, in Eqs.~(84) and~(86) of Ref.~\cite{Aubin:2007mc}.

\section{Synthesis} 
\label{sec:Synthesis}

In this section, we combine our results from the previous two sections, deriving the HQET and \chpt\ descriptions
of heavy-light meson masses, into a practical form.
In addition, we comment on the effect of the nonzero charm quark mass in the relation between the \MSbar\ and MRS masses.

Let us first fix notation for quark masses associated with currently available lattice-QCD ensembles with $2+1+1$ flavors of
quarks~\cite{Bazavov:2012xda}.
We use $m'_l$, $m'_s$ and $m'_c$ to denote the simulation masses of the light, strange and charm quarks, respectively.
We also use $m_l = \frac{1}{2}(m_u+m_d)$, $m_s$ and $m_c$ to denote the tuned values of the corresponding quarks.
As in Sec.~\ref{sec:SChPT}, we use $B_x^{(*)}$ to denote a generic all-staggered heavy-light pseudoscalar
(vector) meson with light valence quark $x$ and heavy valence antiquark $\bar{h}$.

Putting Eqs.~(\ref{eq:match:M0}), (\ref{eq:match:lambda2}), and (\ref{eq:Mass-B-all}) together, using the MRS
mass for the heavy quark, and spelling out the dependence on light quark masses and lattice spacing, we have
\begin{align}
    M_{B_x^{(*)}}(m_x;\{m'_l,m'_l,m'_s\};a) &= \left[ m_{h,\MRS} + \Lambdabar_\MRS 
        + \frac{\mu_\pi^2 - d_{B^{(*)}}\mu_G^2(m_h)}{2m_{h,\MRS}} \right]_\text{chiral-limit}
    \label{eq:fit-function:0}\\
    & + 2\lambda_1 B_0 m_x + 2\lambda'_1 B_0(2m'_l + m'_s) + \delta M_{B_x^{(*)}}(m_x;\{m'_l,m'_l,m'_s\};a) ,
    \nonumber
\end{align}
where the last term is given by the suitable expression from Eqs.~(\ref{eq:dMB}), (\ref{eq:dMB*}), (\ref{eq:dMB-2+1}),
or~(\ref{eq:dMB*-2+1}).
In \eq{fit-function:0}, the HQET matrix elements $\Lambdabar_\MRS$, $\mu_\pi^2$ and $\mu_G(m_h)$ appear
in the SU(3) chiral limit, which is not interesting in practice.
Therefore, we rewrite \eq{fit-function:0} as
\begin{align}
    M_{B_x^{(*)}}(m_x;\{m'_l,m'_l,m'_s\};a) &= m_{h,\MRS} + \Lambdabar_\MRS + \frac{\mu_\pi^2
        - d_{B^{(*)}}\mu_G^2(m_h)}{2m_{h,\MRS}} + 2\lambda_1 B_0 m_x
    \label{eq:fit-function:1} \\ &
        + 2\lambda'_1 B_0(2m'_l + m'_s) + \delta M_{B_x^{(*)}}(m_x;\{m'_l,m'_l,m'_s\};a) - \cC^{(*)} ,
    \nonumber
\end{align} 
where
\begin{equation}
    \cC^{(*)} = 2\lambda_1 B_0 m_q + 2\lambda'_1 B_0(2m_l + m_s) + \delta M_{B_q^{(*)}}(m_q;\{m_l,m_l,m_s\}; 0)
    \label{eq:fit-function:C} 
\end{equation}
is the continuum, physical quark-mass limit of the \chpt\ corrections.
By construction, then, the formula reduces to
\begin{equation}
    M_{B_q^{(*)}}(m_q;\{m_l,m_l,m_s\}; 0) = m_{h,\MRS} + \Lambdabar_\MRS + \frac{\mu_\pi^2
        - d_{B^{(*)}}\mu_G^2(m_h)}{2m_{h,\MRS}}
    \label{eq:M_Hl}
\end{equation}
for $a=0$, tuned sea-quark masses, and $m_q$ takes one of the values $\half(m_u+m_d)$, $m_u$, $m_d$, or $m_s$, depending on the
light flavor of interest.
The resulting values of $\Lambdabar_\MRS$, $\mu_\pi^2$ and $\mu_G^2(m_h)$ are now appropriate for calculations within HQET.

From HQET power counting, it may seem that chiral corrections may be moved arbitrarily between the HQET matrix elements and the
heavy quark mass.
It is crucial, however, that the heavy-quark mass be defined so that it does not depend on the hadron of which it is a constituent.
More formally, the heavy-quark mass should be a singlet under chiral symmetry and appear in all hadron masses in a way consistent
with HQET.
Therefore, it is necessary to use \hmchpt\ to separate all low-energy light-quark mass effects from $M_{B^{(*)}}$ to obtain $M_0$,
and then to use HQET to separate $M_0$ into $m_{h,\MRS}$, $\Lambdabar_\MRS$, and $\mu_\pi^2/2m_Q$.
In this way, all chiral corrections are absorbed into $\Lambdabar_\MRS$, $\mu_\pi^2$, $\mu_G^2$, etc.\ in \eq{M_Hl}.

One can use \eq{mu_G2} to express $\mu_G^2(m_h)$ in terms of $C_\text{cm}(m_h)$ and $\tilde{\mu}_G^2$, which are the Wilson
coefficient and matrix element of the chromomagnetic operator in HQET, respectively.
Equivalently, one can express it as
\begin{equation}
    \mu_G^2(m_h) = \frac{C_\text{cm}(m_h)}{C_\text{cm}(m_b)} \mu_G^2(m_b).
\end{equation}
When lattice-QCD data are not available for the $B^*$ mass (as in Ref.~\cite{Bazavov:2017mbc}), one can use the experimental value
of the hyperfine splitting, $M_{B^*}-M_B$, as a prior on the value of $\mu_G^2(m_b)$.
Because $\mu_G^2(m_b)$ provides the first term in the $1/m_h$ expansion for the physical quantity
$M_{B^*}-M_B$, any renormalon ambiguity must be suppressed by at least another factor of $1/m_h$.
This is not the case for $\mu_\pi^2$, because it is not closely related to any physical quantity, as we discuss briefly below.

A further practical step is to exploit the identity%
\footnote{The left-hand side of Eq.~(\ref{eq:hat}) holds with any mass-independent renormalization scheme,
and the right-hand side relies on the remnant chiral symmetry of staggered fermions.} %
\begin{equation}
    \frac{m_{r,\MSbar}(\mu)}{m_{h,\MSbar}(\mu)} = \frac{am_{r,0}}{am_{h,0}} + \order(a^2),
    \label{eq:hat}
\end{equation}
where $am_0$ is the bare staggered-fermion mass in lattice units, and the subscript $r$ denotes a reference mass.
In Eq.~(\ref{eq:fit-function:1}), we trivially rewrite
\begin{equation}
    m_{h,\MRS} = \frac{m_{r,\MSbar}(\mu)}{m_{h,\MSbar}(\mu)} \frac{am_{h,0}}{am_{r,0}} m_{h,\MRS} =
        m_{r,\MSbar}(\mu) \frac{\mbar_h}{m_{h,\MSbar}(\mu)} \frac{m_{h,\MRS}}{\mbar_h} \frac{am_{h,0}}{am_{r,0}}.
    \label{eq:rabbit}
\end{equation}
The factors in the last expression are, respectively, a convenient fit parameter for lattice-QCD applications, a factor depending on
$\alpha_s$ and the mass anomalous dimension to run between $\mbar_h$ and a fixed scale~$\mu$, an expression given in
Sec.~\ref{sec:MRS-mass}, and a variable that is an input to lattice QCD.
The third factor depends on $\mbar_h$ only through $\ageom(\mbar_h)$, as is easily seen in Eqs.~(\ref{eq:mSI2mMRS}) and
(\ref{eq:J-MRS:def:Incomplete-gamma}).
A different renormalization scheme, besides \MSbar, would amount to substituting the appropriate coefficients for the $r_n$ in
Eq.~(\ref{eq:mSI2mMRS}), and another renormalon subtraction, besides MRS, would amount to substituting the appropriate prescription
for the $\cJ$~function.

Let us now address an important detail in the perturbative series relating the \MSbar\ and MRS definitions of the renormalized quark
mass.
The relation between the $\MSbar$ mass and the pole mass (and hence the MRS mass) is known through order $\alpha_s^4$, but the
order-$\alpha_s^4$ term is known only for massless sea quarks.
The effect of nonzero sea-quark masses is available through order $\alpha_s^3$~\cite{Gray:1990yh,Bekavac:2007tk}.
For the up, down, and strange quarks, the effects of nonzero masses of quarks can be neglected in perturbative QCD.
For the charmed sea quark, however, the nonzero mass does affect the heavy-quark mass at the current precision.
A~bit more generally, let us consider QCD with $n_l$ massless quarks, one charm quark of mass $0<\mbar_c<\mbar_h$, and $n_h$ heavy
quarks of mass $\mbar_h$.
The relation between the pole and $\MSbar$ mass of the $n_h$ heavy quarks is usually first reported with $n_l+1+n_h$ active quarks;
see, for example, Ref.~\cite{Marquard:2016dcn}.
The result, however, is usually expressed and used with $n_l+1$ flavors of active quarks, absorbing the heavy-quark loops into the
running of~$\alpha_s$.
Although the gauge coupling is converted to run with $n_l+1$ active flavors, the heavy-quark mass customarily remains in the
$n_l+1+n_h$ scheme.

When studying the relation between the \MSbar\ and RS masses, \rcite{Ayala:2014yxa}
found a better-behaved expansion if the charm-quark contribution is also decoupled from the renormalon-subtracted series
and added back separately.
This can be explained by the fact that the charm-quark mass cuts off the infrared momenta that generate factorially
growing contributions~\cite{Ball:1995ni}.
In this vein, one can rewrite \eq{mSI2mMRS} as
\begin{equation}
    m_{h,\MRS} = \mbar_h\left(1+\sum_{n=0}^{\infty} \big[r_n^{(n_l)}-R_n^{(n_l)}\big] \alpha_s^{n+1}(\mbar_h; n_l)\right) 
        + \cJ_\MRS^{(n_l)}(\mbar_h) + \Delta m_{(c)},
    \label{eq:mSI2mMRS:decoupled} 
\end{equation}
where $\alpha_s(\mbar_h; n_l)$ is the value of the gauge coupling with $n_l$ active quarks and the charm quark decoupled at scale
$\mu=\mbar_c$; the superscript $(n_l)$ in $r_n^{(n_l)}$, $R_n^{(n_l)}$ and $\cJ_\MRS^{(n_l)}(\mbar_h)$ indicates that they
correspond to the theory with $n_l$ active quarks; and $\Delta m_{(c)}$ includes the contribution of the charmed quark as well as
effects that are generated from matching to the $n_l$-flavor~$\alpha_s$.
More details are provided at the end of this section.
Note that, even though the charmed quark is removed from the perturbative series, the \MSbar\ mass $\mbar_h$ in
Eq.~(\ref{eq:mSI2mMRS:decoupled}) remains in the ($n_l+1+n_h$)-flavor scheme.
In the calculation of the MRS mass, we then set
\begin{equation}
    R_0^{(3)} = 0.535 \pm 0.010,
    \label{eq:renormalon:N}
\end{equation}
which is the overall normalization of the leading infrared renormalon in the pole mass with three massless active
quarks~\cite{Komijani:2017vep}.
The value agrees with other estimates in the literature: $0.563(26)$~\cite{Ayala:2014yxa},
$0.537~\text{(no~error~quoted)}$~\cite{Beneke:2016cbu}, and $0.526(12)$~\cite{Hoang:2017suc}.

To see how these manipulations work in practice, let us now investigate numerically the relation
between the MRS and $\MSbar$ masses for $n_l=3$, $R_0^{(3)}=0.535$, and $n_h=0$
(which is appropriate for 2+1+1-flavor lattice QCD).
Exploiting four-loop calculations in~\rcite{Marquard:2016dcn}, we have
\begin{align}
    r_n^{(3)} &= (0.4244,\   1.0351,\   3.6932,\  17.4358, \ldots),
    \label{eq:rn3} \\
    R_n^{(3)} &= (0.5350,\   1.0691,\   3.5966,\  17.4195, \ldots),  
    \label{eq:Rn3}
\end{align}
for $n=0,1,2,3,\ldots$.
Their differences
\begin{equation}
    r_n^{(3)}-R_n^{(3)} = (-0.1106,\  -0.0340,\  0.0966,\  0.0162, \ldots)
  \label{mSI2MRS:sequence}
\end{equation}
are much smaller; consequently, the series in powers of $\alpha_s$ in \eq{mSI2mMRS:decoupled} can be considered to be a well-behaved
series through order $\alpha_s^4$.
This good behavior indicates that the subleading infrared renormalon is not an immediate problem in computing the MRS mass.
In principle, this renormalon is canceled by a corresponding renormalon in~$\mu_\pi^2$, so Eq.~(\ref{mSI2MRS:sequence}) also
suggests that $\mu_\pi^2$ will be numerically stable in the MRS scheme.
This outcome is not surprising, because the effect of this renormalon is known to be small in
Lorentz-invariant regularization schemes~\cite{Neubert:1996zy}.

Let us return to Eq.~(\ref{eq:mSI2mMRS:decoupled}) and give the explicit expression for $\Delta m_{(c)}$,
which is denoted $\delta m_c$ in \rcite{Ayala:2014yxa}.
The key equation is Eq.~(2.15) in \rcite{Ayala:2014yxa}; in our notation,
\begin{equation}
    \Delta m_{(c)} = \left(\delta m_{(c,+)}^{(1)} + \delta m_{(c,\text{dec})}^{(1)} \right)
        \left[\frac{\alpha_s(\mbar_h; n_l)}{\pi}\right]^2 +
        \left(\delta m_{(c,+)}^{(2)} + \delta m_{(c,\text{dec})}^{(2)} \right)
        \left[\frac{\alpha_s(\mbar_h; n_l)}{\pi}\right]^3,
\end{equation}
where the coefficients $\delta m_{(c,+)}^{(1)}$, $\delta m_{(c,\text{dec})}^{(1)}$, and $\delta m_{(c,\text{dec})}^{(2)}$ are given
in Eqs.~(2.6), (2.16), and~(2.17) of \rcite{Ayala:2014yxa}, and the coefficient $\delta m_{(c,+)}^{(2)}$ is presented in
Ref.~\cite{Bekavac:2007tk}.
Because the quarks that are heavier than the charm quark are (still) quenched in lattice-QCD simulations,
we cannot directly use the expressions provided in \rcite{Ayala:2014yxa}.
As above, we use $n_h$ to denote the number of dynamical heavy quarks with mass $\mbar_h$;
then, for $n_l=3$ and arbitrary $n_h$, one has
\begin{align}
    \delta m_{(c,\text{dec})}^{(2)} = &\, \mbar_h \left[
      -\frac{115981}{11664} -\frac{29 \pi^2}{24} +\frac{61 \pi^4}{1944} 
      -\frac{11}{81} \pi^2 \ln 2 +\frac{2}{81} \pi^2 \ln^2 2 + \frac{\ln^4 2}{81}
    \right . \label{eq:delta-mc-dec-2} \\ & 
      +\frac{8}{27}\,\text{Li}_4(\half)
      -\frac{511}{216} \zeta(3)
      -\ln z^2 \left(
      \frac{377}{96}
      +\frac{\pi^2}{27} \ln 2
      -\frac{1}{18} \zeta(3)
      -\frac{1}{27} \ln z^2 \right)
    \nonumber \\ & \left. \vphantom{\frac{\ln^42}{81}}
      +\frac{n_h}{11664} \left(5917 - 864 \zeta(3) - 468 \pi^2 - 432 \pi^2 \ln z^2 + 3861 \ln z^2 \right)\right]
      -\frac{1}{3} \delta m_{(c,+)}^{(1)}  \ln z^2,
    \nonumber
\end{align}
where $z = \mbar'_c/\mbar_h$, and $\mbar'_c$ is the simulation charmed-quark mass.
To obtain this expression from Ref.~\cite{Ayala:2014yxa}, results in Refs.~\cite{Chetyrkin:1997sg,Melnikov:2000qh} are needed 
to restore the $n_h$ dependence.
Equation~(\ref{eq:delta-mc-dec-2}) reduces to Eq.~(2.17) of \rcite{Ayala:2014yxa} for $n_h=1$.
For analysis of lattice-QCD data with 2+1+1 flavors of sea quarks, one should set $n_h=0$.
Further, to avoid the lengthy expressions for $\delta m_{(c,+)}^{(1)}$ and especially $\delta
m_{(c,+)}^{(2)}$, we use the approximate expressions provided in Eqs.~(3.2) and (3.4) of Ref.~\cite{Bekavac:2007tk}.
We find numerically, for $n_l=3$ and $n_h=0$,
\begin{align}
    \delta m_{(c,+)}^{(1)}          & = \mbar'_c \left(1.596 - 0.6285 z + 0.1777 z^2 \right), \\
    \delta m_{(c,\text{dec})}^{(1)} & = -\mbar_h \left[ 1.0414 + 0.4444 \ln z\right], \\
    \delta m_{(c,+)}^{(2)}          & = \mbar'_c \left[19.987 + 2.824 z + 1.288 z^2 - 
        \left(13.644 + 2.788 z - 0.0343 z^2\right)\ln z\right], \\
    \delta m_{(c,\text{dec})}^{(2)} & = -\mbar_h \left[22.312  + \left(8.227 + 1.064 z - 0.419 z^2 + 0.118 z^3 - 
        0.148\ln z\right)\ln z\right] .
\end{align}
The relations for $\delta m_{(c,+)}^{(2)}$ and $\delta m_{(c,\text{dec})}^{(2)}$ change slightly for $n_h=1$:
\begin{align}
    \delta m_{(c,+)}^{(2)}          & = \mbar'_c \left[19.987 + 3.028 z + 1.141 z^2 - 
        \left(13.644 + 2.788 z - 0.0343 z^2\right)\ln z\right], \\
    \delta m_{(c,\text{dec})}^{(2)} & = -\mbar_h \left[22.29  + \left(8.296 + 1.064 z - 0.419 z^2 + 0.118 z^3 - 
        0.148\ln z\right)\ln z\right],
\end{align}
which may prove useful at a later date.

\section{Summary and Outlook} 
\label{sec:Summary}

As we show in a companion paper~\cite{Bazavov:2017mbc}, the theoretical developments in this paper provide
a new way to determine the quark masses as well as HQET matrix elements such as $\Lambdabar$ and~$\mu_\pi^2$.%
\footnote{The data used in Ref.~\cite{Bazavov:2017mbc} lack the vector-meson masses.
An ideal analysis would include them, in order to obtain information on $\mu_G^2$ as well.} %
With extremely precise lattice-QCD data for the heavy-light pseudoscalar-meson correlation functions~\cite{Bazavov:2017fBD},
Ref.~\cite{Bazavov:2017mbc} yields some of the most precise determinations of all quark masses (other than the top quark).
We find there that the fits of the lattice-QCD data are much more convenient with the MRS mass than with alternatives.
The key is that we do not have to choose a renormalization scale, such as $\nu_f$ for the RS scheme, while preserving the natural
structure of HQET.
We have tried the HQET fits with the RS mass and found that we had to introduce three scales in all, $\nu_f<\mu<\mbar_h$, with $\mu$
being used for~$\alpha_s$~\cite{Komijani:2016jrh}.
In addition, the end results for $\mbar_c$ and $\mbar_b$ depend on $\nu_f$ more than one would like.
These experiences prompted us to reexamine the idea of renormalon subtraction, leading to the MRS scheme.
With the available data, we also found more stable fits to the $m_h$ dependence when using a wide range of data for~$m_x$
and~$m'_l$.
This experience prompted us to derive the \aschpt\ expressions in Sec.~\ref{sec:SChPT}.

The idea to fit the HQET formula, Eq.~(\ref{eq:mQ_2_MH}), to lattice-QCD data to obtain $\bar{\Lambda}$ and~$\mu_\pi^2$ has been
tried before.
The first attempt~\cite{Kronfeld:2000gk} used lattice perturbation theory to convert the bare lattice mass to the pole mass.
In addition to suffering from the renormalon ambiguity, the conversion was (and usually will be) limited to order~$\alpha_s$.
Closer in spirit is the recent work~\cite{Gambino:2017vkx} using nonperturbative renormalization of the lattice quark mass, together
with the kinetic scheme~\cite{Uraltsev:1996rd,Uraltsev:2001ih} in Eq.~(\ref{eq:mQ_2_MH}).
The conversion from the kinetic to the \MSbar\ scheme is known to (only) two loops~\cite{Gambino:2011cq}.
With the MRS scheme, one can use the four-loop relation between the \MSbar\ mass and the pole mass~\cite{Marquard:2016dcn} and, in
the future, any higher-order terms that may be obtained.

For lattice QCD with staggered sea and valence quarks, we augment the MRS scheme with \chpt\ for staggered fermions.
Similar procedures can be applied with other lattice fermion formulations.
For example, \chpt\ for Wilson fermions addresses chiral-symmetry breaking from the dimension-five Wilson term in a way consistent
with that of the mass term.

One can, in principle, extend the definition of the MRS mass by subtracting sequences of higher and higher renormalons from the
perturbative series and collecting their unambiguous contributions into a generalization of the function $\cJ_\MRS(\mbar)$.
The key is to split the Borel integral, as in Eqs.~(\ref{eq:JMRS:def}) and~(\ref{eq:deltam:def}), into unambiguous and ambiguous
pieces, and sweep the ambiguity into $\delta m$.
With some imagination and work, it may be possible to extend the MRS idea beyond renormalon singularities to nonperturbative saddle
points, \ie instantons.
One may then speculate that the remaining series in powers of $\alpha_s$ could acquire a nonzero radius of convergence.
Clearly, such a step will require generalizations of Ref.~\cite{Komijani:2017vep}, or other approaches such as direct investigation
of divergences in the Borel plane, and, thus, remains a conjecture at this stage.

How might this work? %
As explained in Ref.~\cite{Berry:1991XXX}, ``an asymptotic series \ldots\ is a compact encoding of a function, and its divergence
should be regarded not as a deficiency but as a source of information about the [underlying] function.'' %
The crucial point here is that, equipped with the overall normalization~\cite{Komijani:2017vep}, we could decode the divergence due
to the leading infrared renormalon in the pole mass.
Then we could recover the contribution of this renormalon in the form of the nonperturbative function $\cJ_\MRS(\mbar)$, which has a
convergent expansion in powers of the reciprocal of $\alpha_s(\mbar)$, rather than in powers of $\alpha_s(\mbar)$ itself.

The MRS scheme should also be useful in other phenomenological settings---the extra scale(s) of alternative short-distance
masses require more choices, making uncertainty estimation difficult.
For example, the rate of inclusive semileptonic $B$ decay is proportional to $m^5_{b,\pole}$, multiplied by a perturbative series
that cancels the renormalon ambiguities~\cite{Neubert:1994wq,Beneke:1994bc}.
Using instead the MRS scheme for mass and corrections, one would have the product of two unambiguous expressions.

In theoretical treatments of quarkonium, the heavy-quark potential contains a renormalon ambiguity canceled by
$2m_\pole$~\cite{Pineda:1998PhD,Hoang:1998nz}.
An immediate alternative is the MRS scheme, moving the minimal renormalon from the mass to the potential.
A first application could be in the extraction of $\alpha_s$ from the comparison of the static energy computed in
perturbation theory with lattice QCD, as implemented with a different renormalon 
subtraction scheme in Ref.~\cite{Bazavov:2012ka,*Bazavov:2014soa}.
It would be also interesting to use the MRS mass to examine the quarkonium spectrum, for example to determine the quark mass,
either along the lines of Refs.~\cite{Brambilla:2001fw,Kiyo:2013aea,Mateu:2017hlz} or using lattice~QCD.

Another application may be the gluelump mass~\cite{Campbell:1985kp}, which arises in the short-range treatment of
hybrids~\cite{Brambilla:1999xf,Bali:2003jq,Berwein:2015vca}.
As an adjoint-representation analog of $\Lambdabar$, the gluelump mass either contains a renormalon (as in
Ref.~\cite[Eq.~(6)]{Bali:2003jq}) or depends on an artificial scale (as in some other subtraction schemes).
The MRS approach could be extended to provide a physical definition, making possible a direct lattice-QCD calculation along the
lines of our companion determination of quark masses and $\Lambdabar_\MRS$~\cite{Bazavov:2017mbc}.
Similar quantities can be formulated in baryons and tetraquark states with one or more heavy quarks.
Although all examples mentioned here are closely related to the renormalons in the pole mass and-or heavy-quark potential,
renormalons appear in other contexts in QCD, and the MRS scheme may be useful in those areas too.

\acknowledgments

J.K. and A.S.K. thank Martin Beneke, Claude Bernard, and Doug Toussaint for conversations that influenced this paper.
This work was supported in part by the German Excellence Initiative and the European Union Seventh Framework Program under grant
agreement No.~291763 as well as the European Union's Marie Curie COFUND program (J.K., A.S.K.).
The work of N.B. and A.V. was supported  by the DFG cluster of excellence ``Origin and structure of the universe''.
Fermilab is operated by Fermi Research Alliance, LLC, under Contract No.\ DE-AC02-07CH11359 with the United States Department of
Energy, Office of Science, Office of High Energy Physics.
The United States Government retains and the publisher, by accepting the article for publication, acknowledges that the United
States Government retains a non-exclusive, paid-up, irrevocable, world-wide license to publish or reproduce the published form of
this manuscript, or allow others to do so, for United States Government purposes.


\bibliographystyle{apsrev4-1}
\bibliography{References.bib}

\end{document}